\documentclass[11pt, submission, a4paper, Proceedings]{SciPost}
\pdfoutput=1

\hypersetup{
    colorlinks,
    linkcolor={red!50!black},
    citecolor={blue!50!black},
    urlcolor={blue!80!black}
}

\usepackage[bitstream-charter]{mathdesign}
\usepackage[symbol]{footmisc}
\usepackage[separate-uncertainty=true]{siunitx}
\usepackage{cleveref}
\usepackage{subcaption}
\usepackage[]{graphicx}
\DeclareSymbolFont{usualmathcal}{OMS}{cmsy}{m}{n}
\DeclareSymbolFontAlphabet{\mathcal}{usualmathcal}

\begin{document}

\begin{center}{\Large \textbf{
Neutrinos from near and far: Results from the IceCube Neutrino Observatory\\
}}\end{center}

\begin{center}
Tianlu Yuan\textsuperscript{1$\star$} for the IceCube Collaboration\footnote[2]{\protect\url{https://icecube.wisc.edu}}
\end{center}

\begin{center}
{\bf 1} Dept. of Physics and Wisconsin IceCube Particle Astrophysics Center, University of Wisconsin–Madison, Madison, WI 53706, USA
\\
* tyuan@icecube.wisc.edu
\end{center}

\begin{center}
\today
\end{center}


\definecolor{palegray}{gray}{0.95}
\begin{center}
\colorbox{palegray}{
  \begin{tabular}{rr}
  \begin{minipage}{0.1\textwidth}
    \includegraphics[width=30mm]{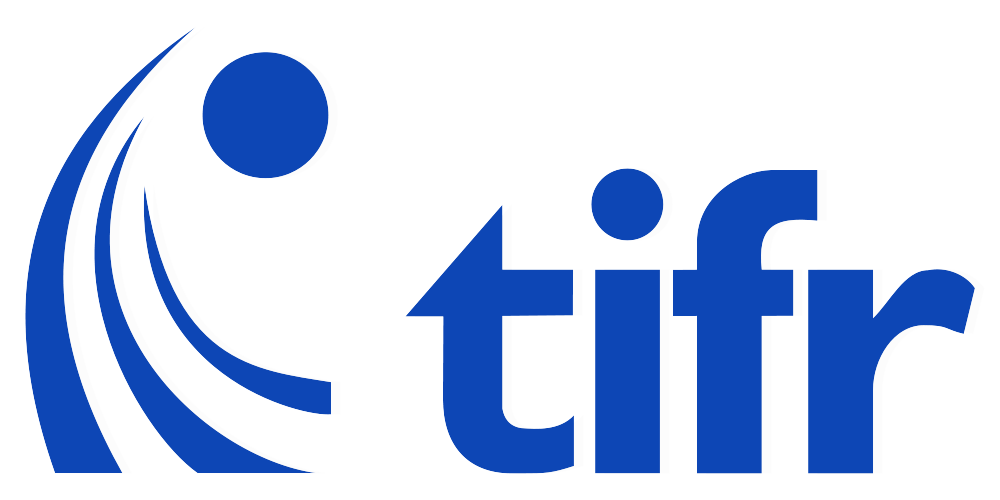}
  \end{minipage}
  &
  \begin{minipage}{0.85\textwidth}
    \begin{center}
    {\it 21st International Symposium on Very High Energy Cosmic Ray Interactions (ISVHE- CRI 2022)}\\
    {\it Online, 23-27 May 2022} \\
    \doi{10.21468/SciPostPhysProc.?}\\
    \end{center}
  \end{minipage}
\end{tabular}
}
\end{center}

\section*{Abstract}
{\bf
Instrumenting a gigaton of ice at the geographic South Pole, the IceCube Neutrino Observatory has been at the forefront of groundbreaking scientific discoveries over the past decade. These include the observation of a flux of TeV-PeV astrophysical neutrinos, detection of the first astrophysical neutrino on the Glashow resonance and evidence of the blazar TXS 0506+056 as the first known astronomical source of high-energy neutrinos. Several questions, however, remain, pertaining to the precise origins of astrophysical neutrinos, their production mechanisms at the source and in Earth’s atmosphere and in the context of physics beyond the Standard Model. This proceeding highlights some of our latest results, from new constraints on neutrino interactions and oscillations to the latest measurements of the astrophysical neutrino flux and searches for their origins to future prospects with IceCube-Gen2.
}

\vspace{10pt}
\noindent\rule{\textwidth}{1pt}
\tableofcontents\thispagestyle{fancy}
\noindent\rule{\textwidth}{1pt}
\vspace{10pt}

\section{Introduction}
\label{sec:intro}
\begin{figure}[hbt]
\centering
\includegraphics[width=0.54\columnwidth]{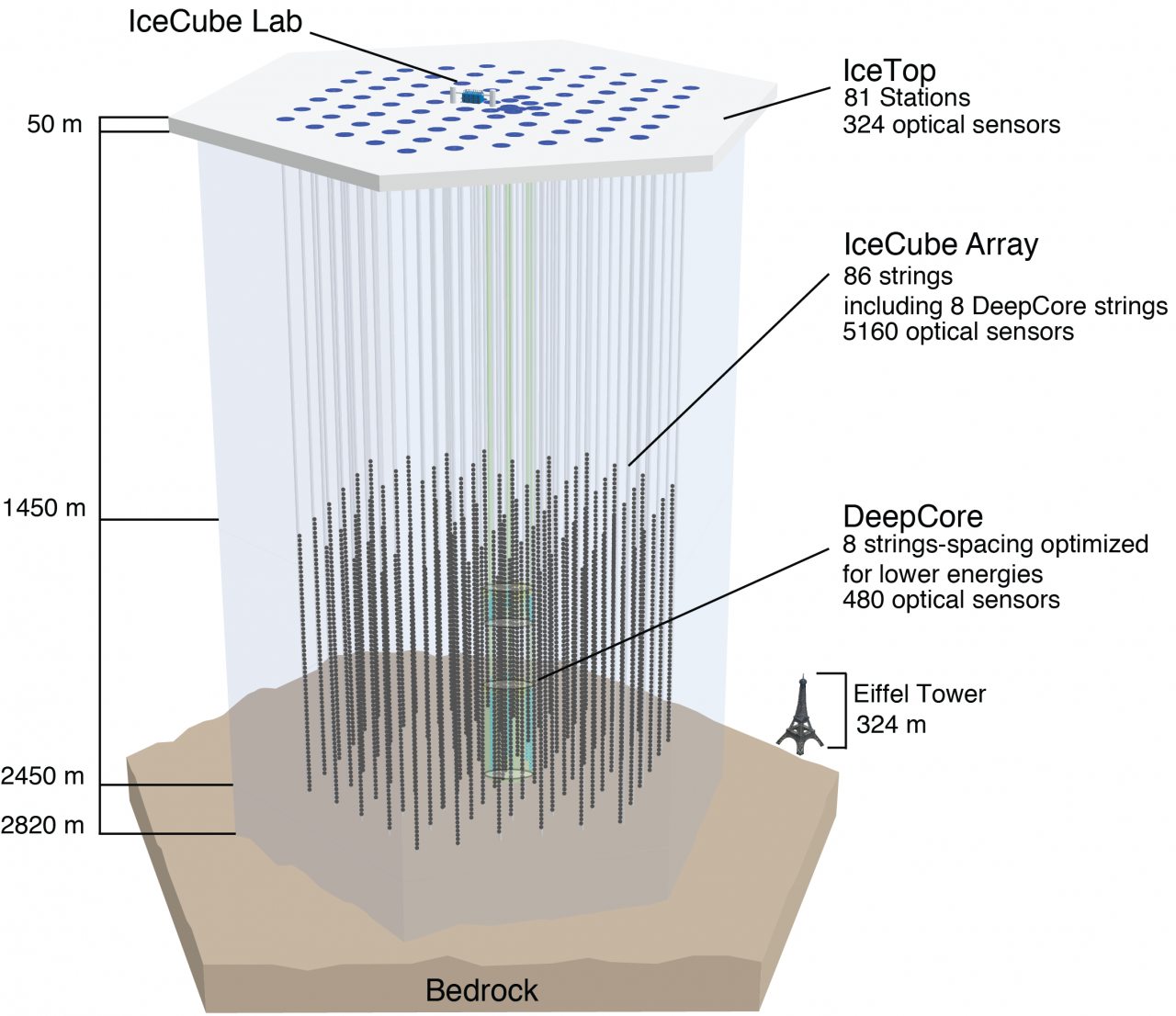}
\includegraphics[width=0.41\columnwidth]{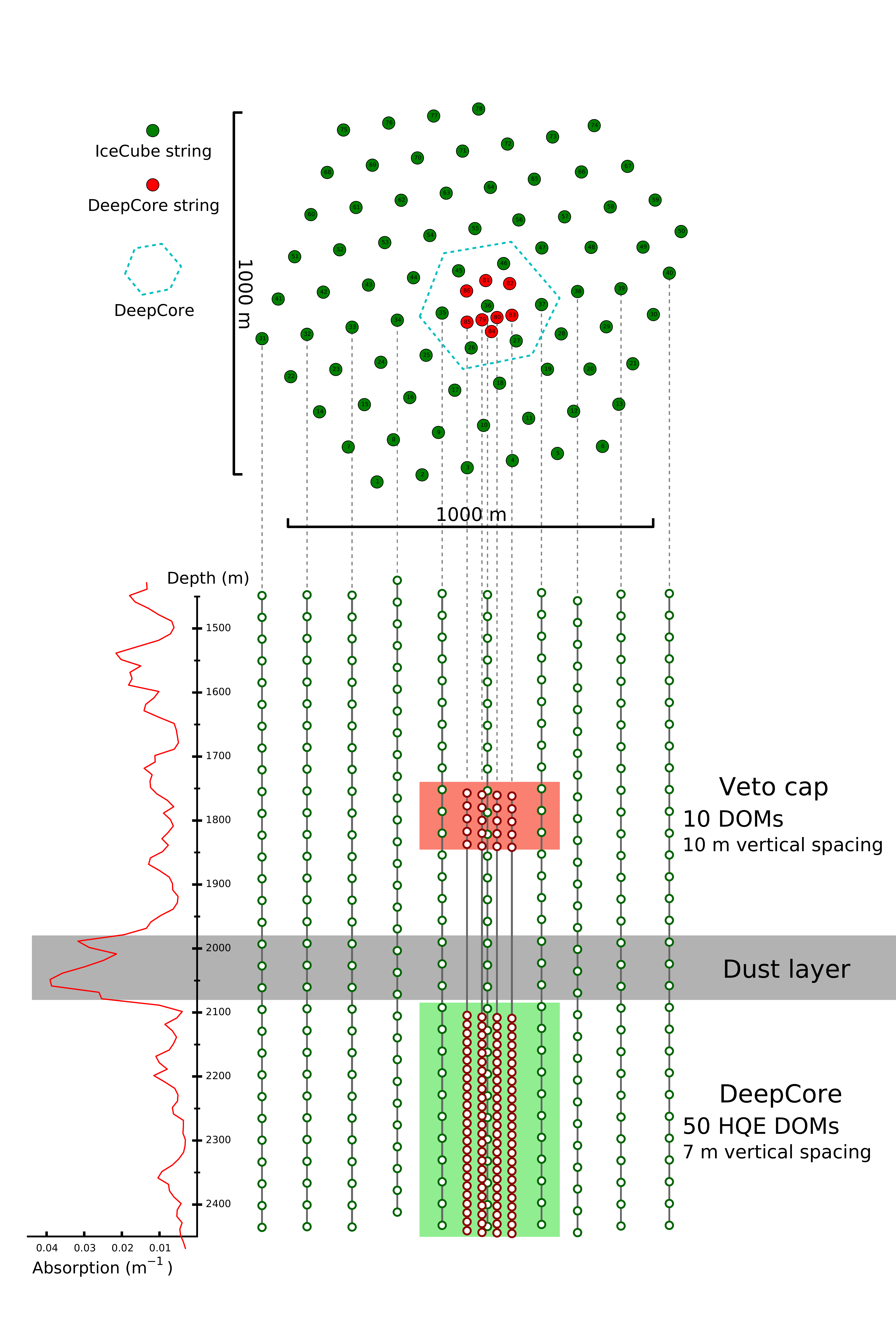}
\caption{Left: The IceCube Neutrino Observatory~\cite{Aartsen:2016nxy}. DeepCore is highlighted in the center of the detector. Right: The red curve shows the absorption coefficient as a function of depth along a side-view of the detector~\cite{Aartsen:2019tjl}. A layer of heightened absorption, called the dust layer, is highlighted in grey.}
\label{fig:detector}
\end{figure}

The IceCube Neutrino Observatory detects neutrinos interacting with nucleons and electrons in the South Pole ice via Cherenkov radiation produced by charged secondaries. It is instrumented with 5160 Digital Optical Modules (DOM), each with a single downward-facing photomultiplier tube (PMT), arrayed across a cubic kilometer~\cite{Aartsen:2016nxy}. The DOMs are attached to 86 strings --- cables drilled into the ice that provide mechanical and electrical support. DOMs are spaced \SI{17}{\m} apart on standard IceCube strings and \SI{7}{\m} apart on DeepCore strings, a denser infill region of the detector. Standard IceCube strings are spaced approximately \SI{125}{\m} apart. \Cref{fig:detector} illustrates the scale and hexagonal configuration of the in-ice detector (left panel) as well as the absorption versus depth along the detector (right panel).

Dependent on the interaction channel, most neutrino-induced events in IceCube can be classified into three categories~\cite{IceCube:2020wum}: particle showers, or cascades, from a high-energy electron produced in a charged current (CC) $\nu_e$ interaction or a hadronic shower from the breakup of the nucleon in both CC and neutral current (NC) interactions, muon tracks induced by a CC $\nu_\mu$ interaction that travel linearly through a significant portion of the detector, or a double cascade whereby a CC $\nu_\tau$ interaction produces a $\tau$ that decays after a distinctly separable distance ($\sim \SI{50}{\m \per \peta \eV}$). Other unique detector signatures are possible for example at the Glashow resonance~\cite{IceCube:2021rpz} and in the individual PMT waveforms~\cite{IceCube:2015vkp}. This proceeding highlights the broad physics reach of the detector, including neutrino source searches (\Cref{sec:sources}), diffuse astrophysical flux measurements (\Cref{sec:diffuse}), and particle physics (\Cref{sec:pp}). In addition, prospects for a next generation detector ten times the size of IceCube will be discussed in \Cref{sec:gen2}.


\section{IceCube results}
\label{sec:results}
Depending on the physics of interest, IceCube takes distinct approaches in the analysis of its data. Angular resolution is important in neutrino source searches, which typically look for clustering, correlations, or time dependence between neutrinos or other astroparticles. An unbinned likelihood ratio method is used to maximize sensitivity~\cite{Braun:2008bg}. In measurements of the diffuse flux expected event rates are computed from large-scale Monte Carlo (MC) simulations which can be later scaled to match model predictions. Model parameters can then be fitted under a binned likelihood assumption in the observable space. A similar approach can be taken for measurements of particle physics parameters and searches for exotic particles.

\subsection{Source searches}
\label{sec:sources}
\begin{figure}[htb]
\centering
\includegraphics[width=0.7\columnwidth]{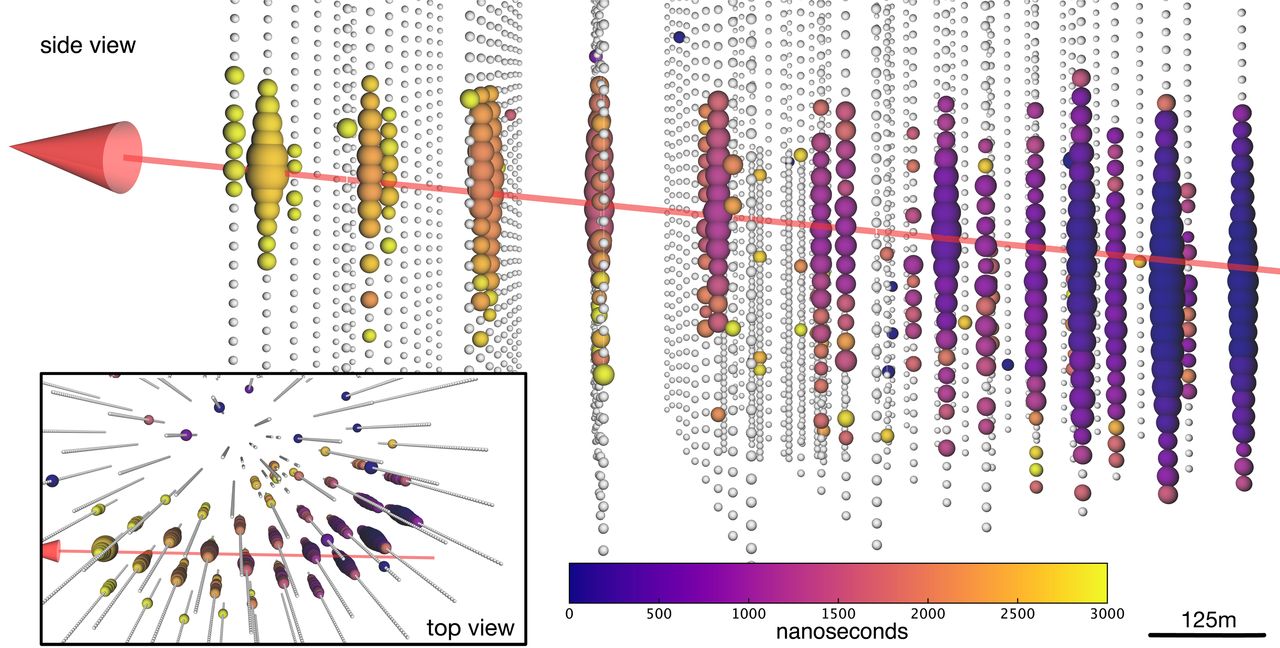}
\caption{Event view of IC170922~\cite{IceCube:2018dnn}. The color of the bubbles indicates the first photon time of arrival and their size corresponds to the total number of photoelectrons detected by the DOM. The red arrow indicates the best-fit reconstructed direction of the track. This event triggered a multimessenger follow-up campaign that led to the discovery of blazars as a source of high-energy astrophysical neutrinos. It also nicely illustrates the operational principles of IceCube.}
\label{fig:txs}
\end{figure}

The relatively accurate directional pointing (\ang{1} or less median angular resolution) of muon tracks makes them the predominant signal in the search for astrophysical neutrino sources. Coupled to the realtime program~\cite{IceCube:2016cqr}, which alerts our multimessenger partners for follow-up observations, IceCube was able to pinpoint TXS 0506+056 as the first candidate source of high-energy astrophysical neutrinos~\cite{IceCube:2018dnn}. An event view of IC170922, a \SI{290}{\tera \eV} track that occurred on September 22, 2017 and triggered the realtime alert, is shown in \Cref{fig:txs}. Shortly after, \textit{Fermi} and MAGIC observed that the blazar TXS 0506+056 was in a flaring state and consistent with the direction of the IceCube track. The chance coincidence probability, including trials correction for previous alerts, was calculated to be disfavored at the $3\sigma$ level. An analysis of archival IceCube data led to the discovery of an excess of neutrino events in 2014-2015 at the location of the blazar~\cite{IceCube:2018cha}. The excess of events consists of lower energy neutrinos, clustered in a 110-day window at a significance of $3.5\sigma$. No excess was observed around the time of the 2017 alert. These two independent analyses -- the coincidence of a high-energy track with a flaring blazar and the archival neutrino ``flare'' -- are compelling evidence that TXS 0506+056 is a neutrino source.

\begin{figure}[htb]
\centering
\includegraphics[width=0.5\columnwidth]{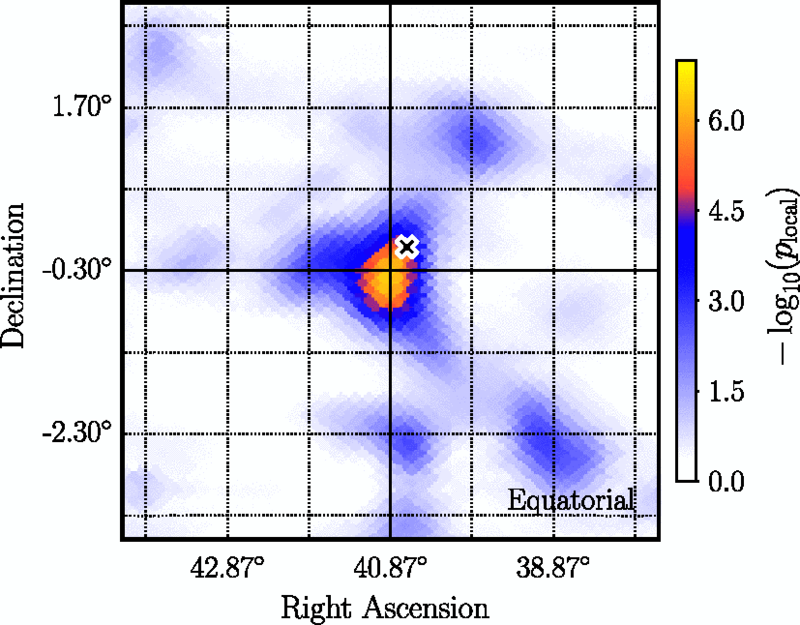}
\caption{The most significant region derived from the all-sky scan located near NGC 1068 ($\times$)~\cite{IceCube:2019cia}. The color map indicates the local significance from the all-sky analysis.}
\label{fig:ngc}
\end{figure}

IceCube has also performed all-sky, time-integrated neutrino source searches. For throughgoing tracks, those that traverse across the detector and typically confer the best directional information, IceCube has the highest sensitivity near the horizon and in the northern sky. The large background of atmospheric muons limits sensitivity in the southern sky. In the most recent publication with ten years of data, three analyses were defined a priori: a full-sky scan, a catalog search, and a stacked search~\cite{IceCube:2019cia}. The most significant result was obtained with the catalog search, where an excess over background was detected coincident with the Seyfert II galaxy NGC 1068 at a post-trial significance of $2.9\sigma$. The all-sky scan is subject to a large trials factor and yielded a post-trial $p=0.099$, while the stacked search yielded $p>0.01$. The catalog search is performed across a list of 110 sources (97 in the northern sky, 13 in the southern sky). Its most significant source, which also corresponds to the hottest spot in the northern sky as shown in \Cref{fig:ngc}, is NGC 1068 at $2.9\sigma$.

Astrophysical neutrinos are likely coupled to ultra-high-energy cosmic rays (UHECR) by their common production mechanisms. In collaboration with ANTARES, Auger, and Telescope Array, IceCube performed a search for correlations between neutrinos and cosmic rays~\cite{ANTARES:2022pdr}. Three analyses were performed, searching for neutrinos inline with UHECR, searching for UHECR inline with neutrinos, and a two-point correlation analysis. While no significant correlation was detected, upper limits were placed on the flux of neutrinos along the direction of UHECRs.

\subsection{Diffuse flux of astrophysical neutrinos}
\label{sec:diffuse}
\begin{figure}[htb]
\centering
\includegraphics[width=0.7\columnwidth]{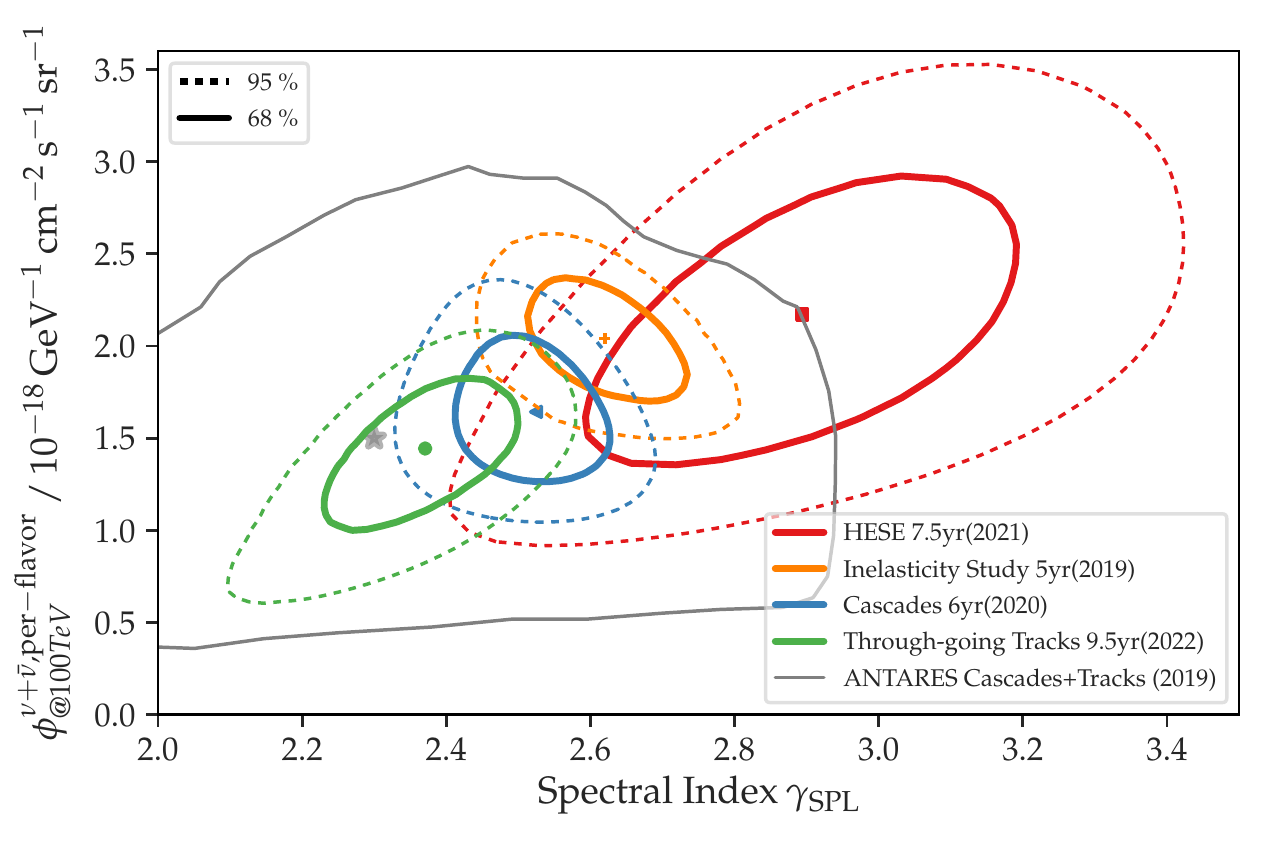}
\caption{Overview of astrophysical diffuse flux measurements to date~\cite{Ackermann:2022rqc}. The green point and contours shows the most recent best-fit and uncertainties using a sample of predominantly up-going tracks in IceCube with 9.5 years of data taking~\cite{IceCube:2021uhz}. It is overlaid with IceCube results using high-energy starting tracks and cascades (red)~\cite{IceCube:2020wum}, predominantly contained cascades (blue)~\cite{IceCube:2020acn}, and a sample of predominantly starting tracks (orange)~\cite{IceCube:2018pgc}. A recent ANTARES result (gray) using both cascades and tracks with 9 years of data is included for comparison~\cite{Fusco:2020owb}.}
\label{fig:spl}
\end{figure}

IceCube has measured the diffuse flux of astrophysical neutrinos using samples that comprise primarily of tracks~\cite{IceCube:2015qii,IceCube:2016umi,IceCube:2021uhz}, cascades~\cite{IceCube:2020acn}, and a mixture of both~\cite{IceCube:2014stg,IceCube:2020wum,IceCube:2018pgc}. For upgoing tracks, the only method to distinguish diffuse astrophysical neutrinos from atmospheric neutrinos is to employ model constraints of their distinct energy spectra. For downgoing events that start within a fiducial region of IceCube, a detector veto is typically required to reject the atmospheric muon background. Such a veto also allows rejection of atmospheric neutrinos, based on the expectation that down-going atmospheric neutrinos can be tagged if accompanied by muons produced in the same air shower~\cite{Schonert:2008is,Gaisser:2014bja,Arguelles:2018awr}. Both approaches have yielded discoveries of an astrophysical flux at TeV-to-PeV energies across the different samples. \Cref{fig:spl} has been adapted from~\cite{IceCube:2021uhz} and shows the current global picture of the astrophysical spectrum, assuming a single-power-law (SPL) flux. All results are consistent at the $2\sigma$ level.

\begin{figure}[htb]
\centering
\includegraphics[width=0.5\columnwidth]{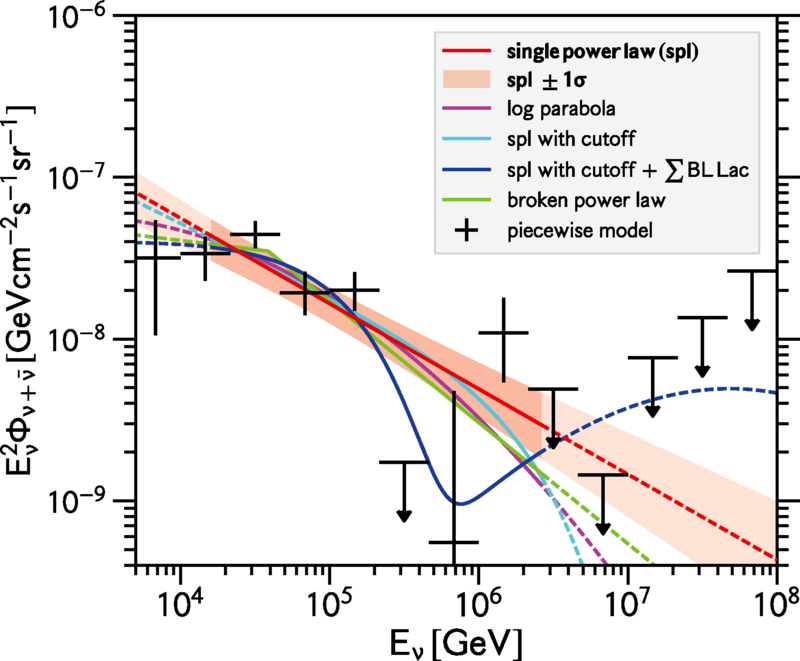}
\caption{Figure 3 from~\cite{IceCube:2020acn} showing the per-flavor astrophysical neutrino flux for the single-power-law (red) and other models. A piecewise measurement is shown in black, while uncertainties on the single-power-law model are shown as the red shaded region. The data is consistent with a single-power-law.}
\label{fig:cascades}
\end{figure}

While the SPL model is well motivated and simple, additional model comparisons have been performed under more sophisticated spectral assumptions. These include tests for a spectral cutoff, a broken power-law, and log-parabola fluxes. The latest result using 9.5 years of upgoing tracks~\cite{IceCube:2021uhz} (\Cref{fig:spl} green) found its data consistent with a SPL hypothesis, but sees hints of softening above \SI{1}{\peta \eV} at a $2\sigma$ level. Inline with the upgoing-track results, \Cref{fig:cascades} shows model comparisons using six years (2010-2015) of cascade data~\cite{IceCube:2020acn} with no significant rejection of the SPL model found. Compared to tracks, cascades typically have a much better energy resolution at the expense of a worse angular resolution. Measurements of the diffuse flux can benefit substantially from improved energy reconstruction, while still employing the atmospheric neutrino self-veto in the downgoing region. The analysis relies on a boosted decision tree (BDT) to select for primarily $\nu_e$ and $\nu_\tau$ CC interactions, with a subset of events expected from NC contributions. The specific model tests are unable to reject the SPL hypothesis, though they indicate a spectral softening at high energies at the $2\sigma$ level. In addition to fits for particular functional form of the flux, a piecewise differential measurement was performed whereby the spectral shape can be probed in a more model-independent manner.

The highest-energy, contained cascade was detected at an energy of \SI{2}{\peta \eV}, below the energy of the Glashow resonance. This resonance is an enhancement of the $s$-channel neutrino charged-lepton cross section, and due to the preponderance of electrons in matter it occurs only for electron antineutrinos on Earth~\cite{Glashow:1960zz}. In the electron rest frame $E_R=\SI{6.3}{\peta \eV}$. Due to the increased cross section, it is strongly suppressed by Earth absorption for neutrinos from the northern sky as illustrated in the left panel of \Cref{fig:xs}, which shows the expected arrival-to-surface flux ratio~\cite{Vincent:2017svp}. To increase the possibility of detecting events on resonance IceCube performed a search for partially contained events, expanding outward from the fiducial volume of the previous analyses~\cite{IceCube:2021rpz}. Without the outer layer of the detector as a veto, stringent BDT-based cuts were applied to reject the large downgoing muon background. Sixty years after its initial proposal, IceCube detected an astrophysical neutrino interacting at the resonance energy for the first time. Its visible energy was reconstructed as $\SI{6.05\pm0.72}{\peta \eV}$, consistent with on-resonance production at $2.3\sigma$ significance. The detection opens an additional identification channel of both the neutrino flavor and charge. As a result, sources of high-energy astrophysical neutrinos can be expected to produce both neutrinos and antineutrinos. Even with only one event detected thus far, the diffuse neutrino flux is now expected to extend to the resonance energy.

\subsection{Particle physics}
\label{sec:pp}
\begin{figure}[htb]
\centering
\includegraphics[width=0.54\columnwidth]{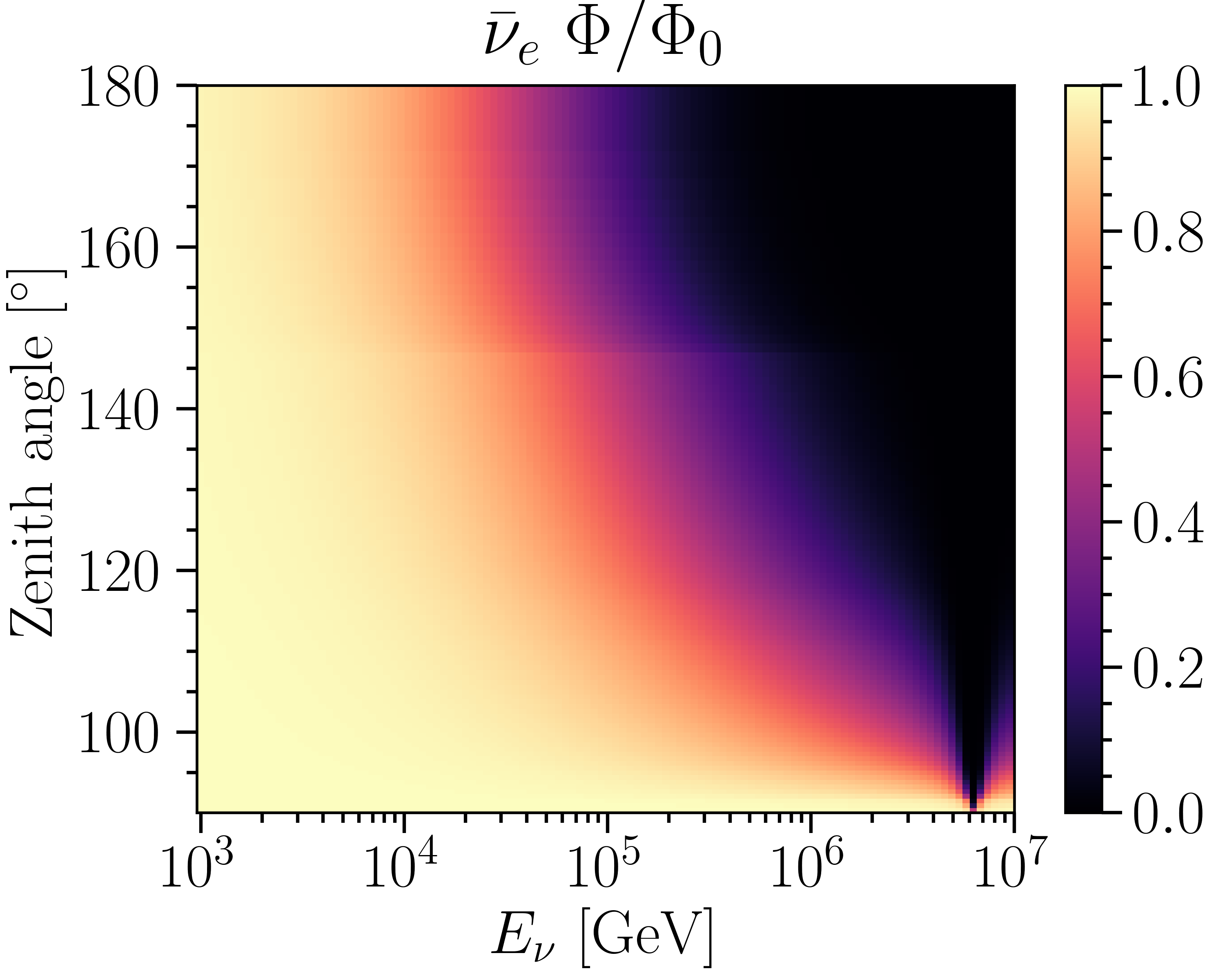}
\includegraphics[width=0.41\columnwidth]{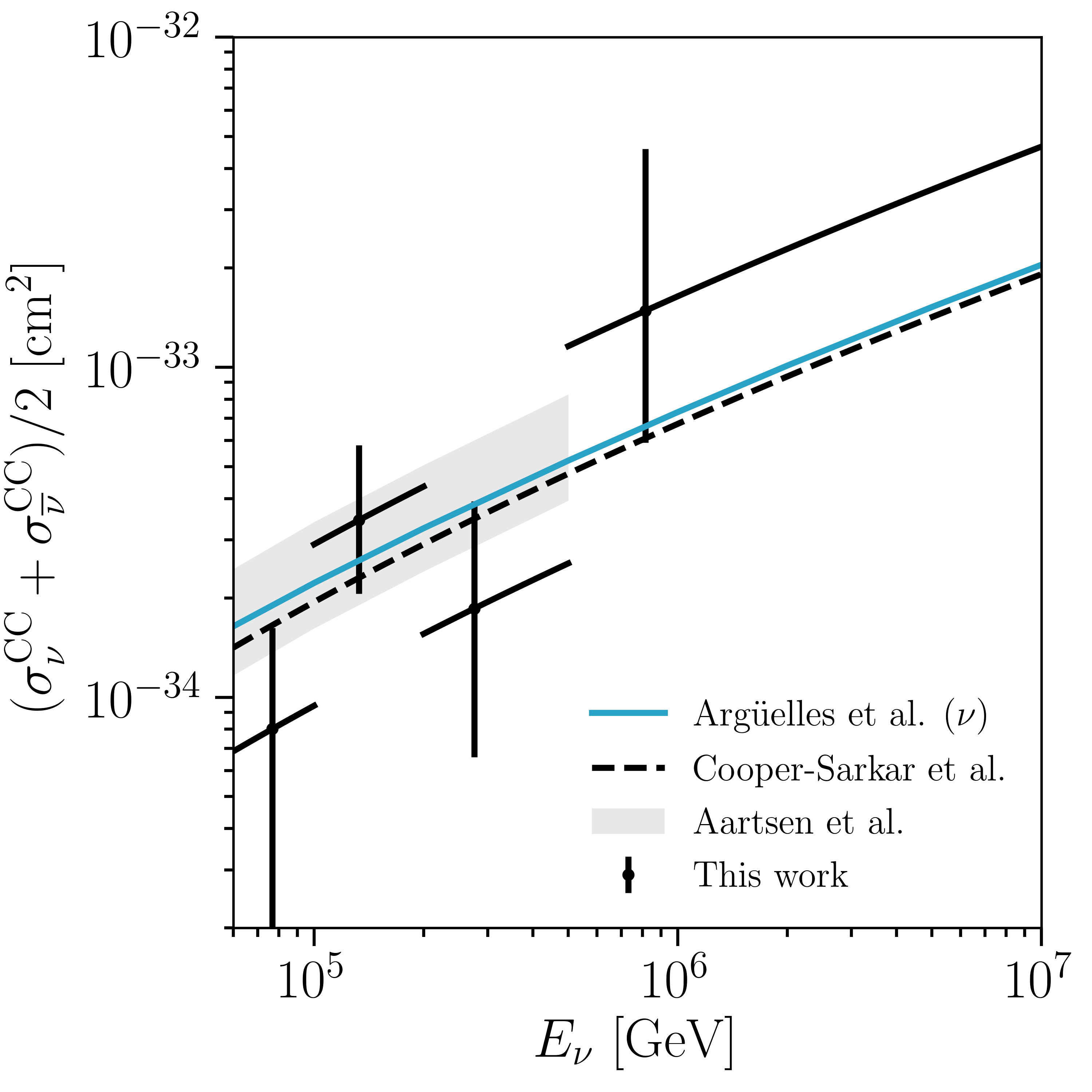}
\caption{Figures from~\cite{IceCube:2020rnc}. Left: arrival-to-surface flux ratio $\Phi/\Phi_0$ for electron antineutrinos for a range of zenith angles and $E_\nu$. The effect of the Glashow resonance is visible at \SI{6.3}{\peta \eV}. Right: measured DIS cross section in four different energy bins (black) using starting events, and total DIS cross section as measured using upgoing tracks (gray). Two Standard Model predictions are shown as blue and black-dashed lines.}
\label{fig:xs}
\end{figure}
As the discussion of the Glashow resonance already alluded to, IceCube is capable of producing exciting results that span multiple subfields of physics. Neutrinos serve not only as astrophysical messengers sent to us from the largest scales in the universe, but also as unique probes of fundamental physics at the smallest scales. These include measurements of the neutrino cross section~\cite{Aartsen:2017kpd,IceCube:2020rnc}, searches for sterile neutrinos~\cite{IceCube:2020phf,IceCube:2020tka} and searches for relativistic magnetic monopoles~\cite{IceCube:2021eye}.

The Earth attenuation of neutrinos starting at roughly \SI{10}{\tera \eV} allows IceCube to measure the neutrino-nucleon deep-inelastic-scattering (DIS) cross section. A modification in the cross section affects the arrival flux at IceCube, thus the data can be used to constrain the cross section in turn. \Cref{fig:xs} illustrates this effect under Standard Model predictions of electron antineutrino cross section~\cite{Cooper-Sarkar:2011jtt}. The color map indicates the ratio of the arrival flux, $\Phi$, to surface flux, $\Phi_0$, over a range of zenith angles and $E_\nu$. A zenith angle of \ang{90} (\ang{180}) corresponds to a path horizontally (diametrically) through the Earth to IceCube. Using a sample of high-energy starting events~\cite{IceCube:2020wum}, an all-flavor measurement of the neutrino-nucleon DIS cross section is measured. In the analysis, the full zenith range from \ang{0} to \ang{180} is used. The result is shown in the right panel of \Cref{fig:xs} and is consistent with the Standard Model predictions. An earlier IceCube measurement based on upgoing tracks is shown as the shaded gray region~\cite{Aartsen:2017kpd}.

\begin{figure}[htb]
\centering
\includegraphics[width=0.4\columnwidth]{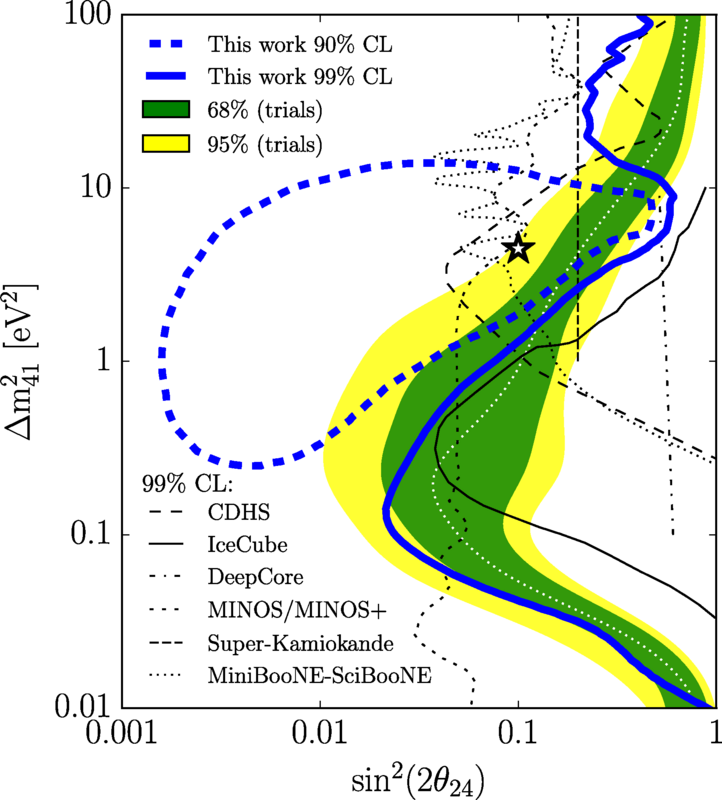}
\includegraphics[width=0.56\columnwidth]{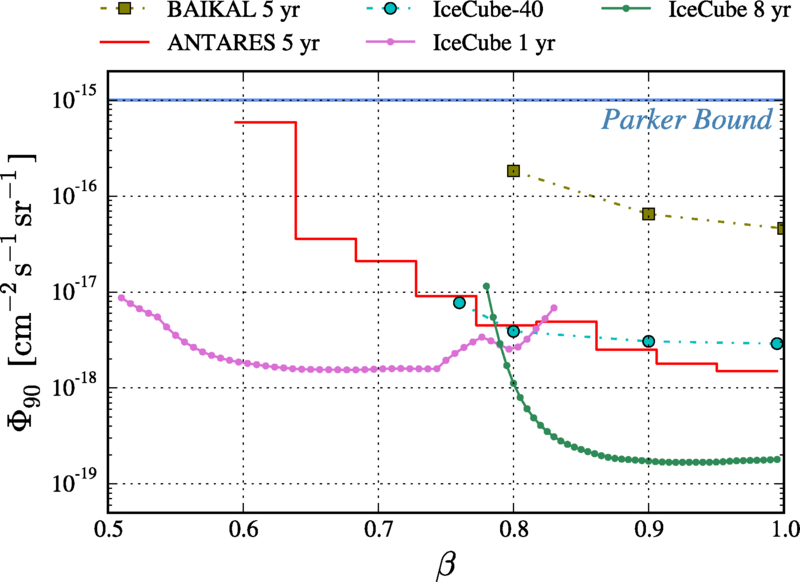}
\caption{Left: figure from~\cite{IceCube:2020phf} that shows the latest IceCube constraints on light sterile neutrino mixing. Right: limits on the flux of magnetic monopoles as a function of velocity $\beta$~\cite{IceCube:2021eye}.}
\label{fig:bsm}
\end{figure}

IceCube has performed dedicated analyses to search for physics beyond the Standard Model (BSM). In particular, IceCube has searched for sterile neutrinos outside the three-flavor neutrino oscillation paradigm~\cite{IceCube:2020phf} and placed constraints on the $3+1$ parameters $\Delta m^2_{41}$-$\sin ^2(2\theta_{24})$ as shown in the left panel of \Cref{fig:bsm}. The best-fit (star) lies at $\sin ^2(2\theta_{24})=0.10$ and $\Delta m^2_{41} =\SI{4.5}{\eV^2}$ and is consistent with three flavor oscillations at $p=0.08$. The analysis uses eight years of upgoing muon track data, spanning a reconstructed muon energy range from \SIrange{500}{9976}{\giga \eV}. 

In the search for more exotic particles, IceCube recently placed the most stringent upper limit on the flux of relativistic magnetic monopoles at $\beta > 0.8$~\cite{IceCube:2021eye}. Such particles are expected to produce a slowly propagating track with uniform light deposition along its length. Zero events were detected that passed all selection cuts, allowing IceCube to place upper bounds on the monopole flux as shown in the right panel of~\Cref{fig:bsm}.

\section{IceCube-Gen2}
\label{sec:gen2}
\begin{figure}[htb]
  \centering
  \begin{minipage}[t]{.78\linewidth}
    \subcaptionbox{IceCube-Gen2 schematic}
      {\includegraphics[width=\linewidth]{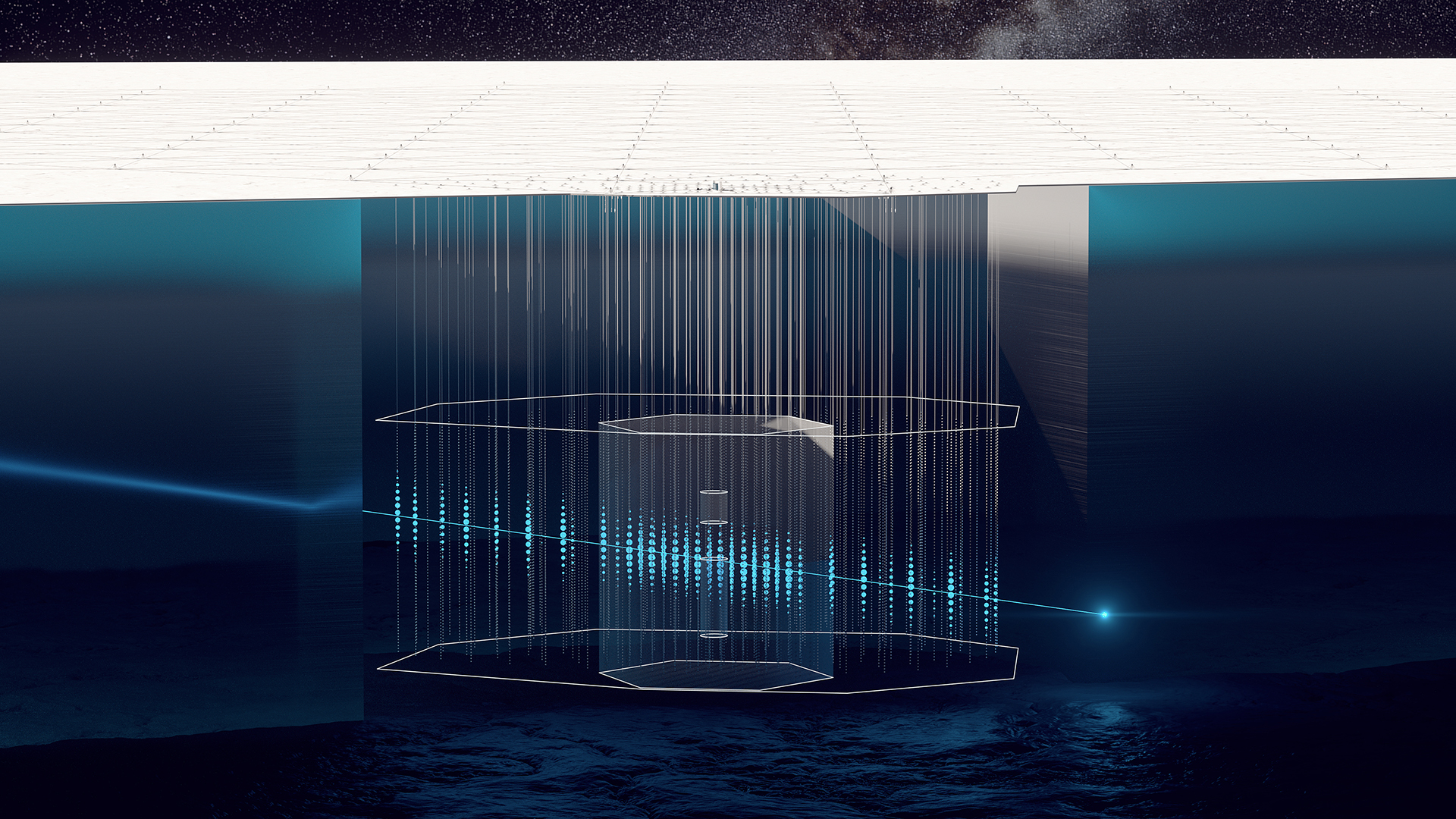}}%
  \end{minipage}%
  \hfill
  \begin{minipage}[b]{.15\linewidth}
    \subcaptionbox{16 PMTs}
      {\includegraphics[width=\linewidth]{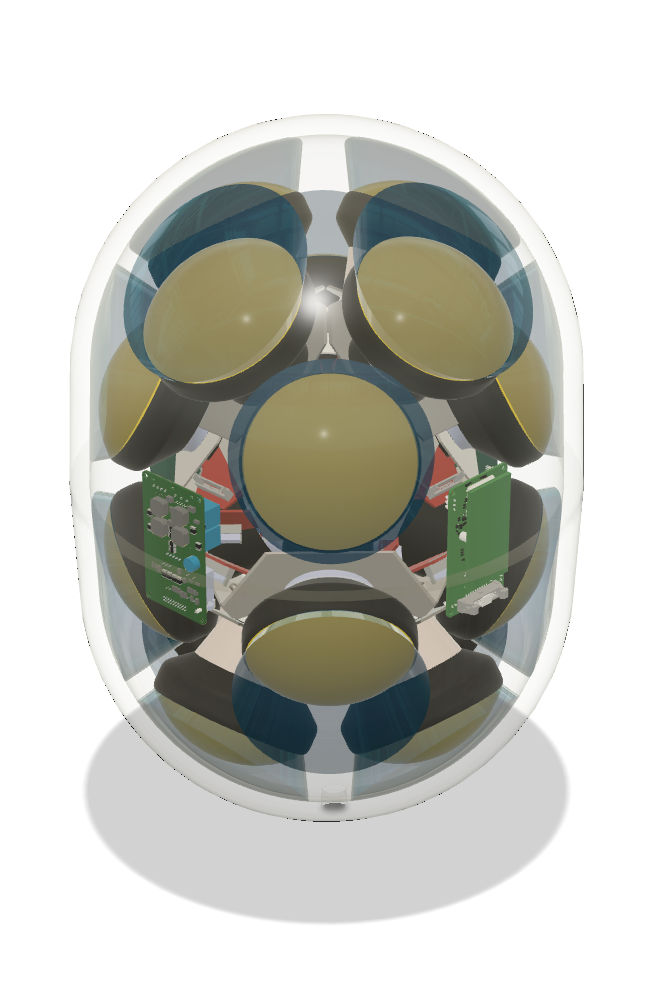}}
    \subcaptionbox{18 PMTs}
      {\includegraphics[width=\linewidth]{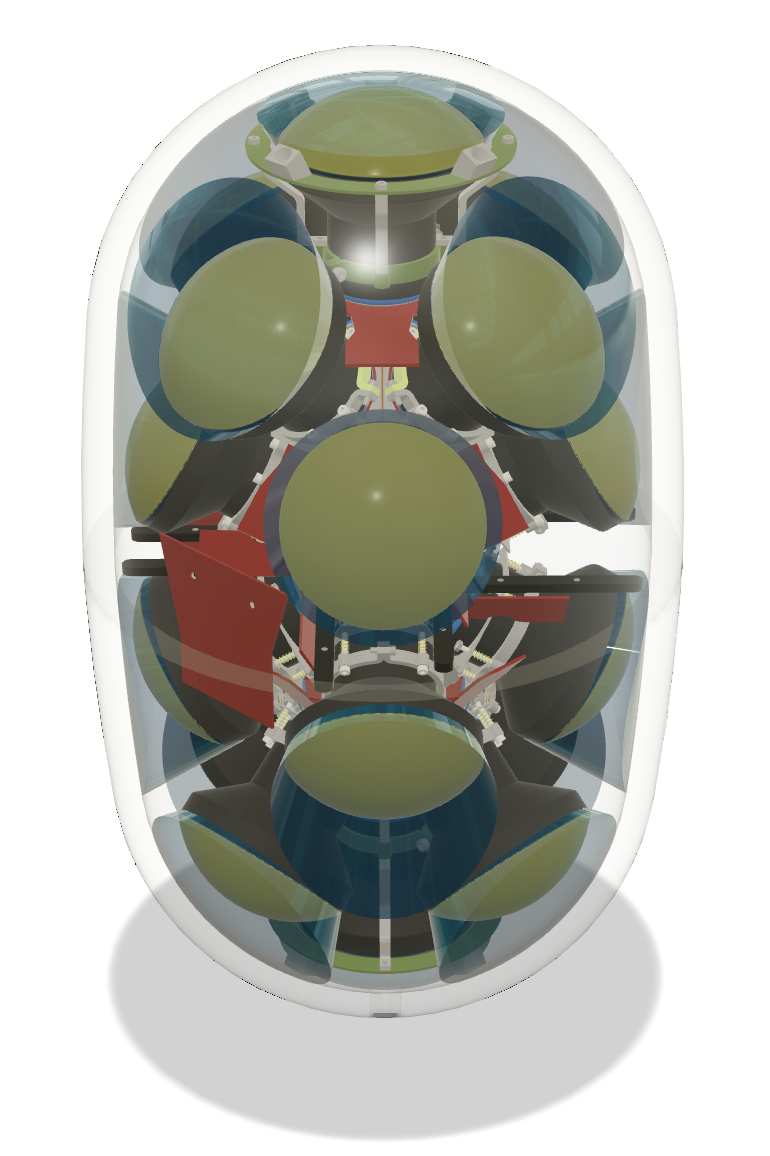}}%
  \end{minipage}%
  \caption
    {%
      The envisioned IceCube-Gen2 facility (a) with in-ice optical, surface, and radio arrays shown. The current IceCube detector is inset. Two prototype sensors for the Gen2 optical array are also shown: a 16 PMT configuration (b) and an 18 PMT configuration (c)~\cite{Makino:lom}.%
    }%
\label{fig:gen2}%
\end{figure}

Building on the success of IceCube, the next generation in-ice neutrino telescope is under development~\cite{IceCube-Gen2:2020qha}. It will comprise an in-ice optical array with 120 additional strings that cover ten times the volume of IceCube. The new optical sensors will increase photocathode coverage using multiple PMTs, thus capturing additional directional information. Two prototypes designs are shown in \Cref{fig:gen2} (right). In addition to the optical array, IceCube-Gen2 will include a surface array~\cite{IceCube-Gen2:2021aek} for cosmic-ray physics and a sparse radio array covering \SI{500}{\km^2} for EeV neutrino detection~\cite{IceCube-Gen2:2021rkf}. Much as IceCube transformed the upper bounds measured by its predecessor AMANDA into measured fluxes, the full IceCube-Gen2 configuration will extend sensitivity above \SI{10}{\peta \eV} and its improved angular resolution for tracks will usher in a new era of precision neutrino astronomy.

\section{Conclusion}
\label{sec:conclusion}
The IceCube Neutrino Observatory continues to produce scientific discoveries in astrophysics and particle physics. Recent results improve on previous measurements while yielding new discoveries. Its excellent uptime and broad energy range ensures that it remains still a one of a kind instrument.

\section*{Acknowledgements}
\paragraph{Funding information}
TY is supported by NSF grant PHY-1913607.



\bibliography{main.bib}

\nolinenumbers

\end{document}